\documentclass[useAMS,usenatbib]{mn2e}
\usepackage{epsfig}
\usepackage{amssymb}
\usepackage{amsmath}

\title[Selection of ULIRGs in Infrared and Submm Surveys]{Selection of ULIRGs in Infrared and Submm Surveys}
\author[Symeonidis et al.]{M. Symeonidis$^{1}$\thanks{msy@mssl.ucl.ac.uk}, M. J. Page$^{1}$ and N. Seymour$^{1}$\\
$^{1}$University College London, Mullard Space Science Laboratory,
Holmbury St. Mary, Dorking, Surrey RH5 6NT, UK.}

\begin{document}

\date{Accepted  Received; in original form}

\pagerange{\pageref{firstpage}--\pageref{lastpage}} \pubyear{2010}

\maketitle

\label{firstpage}

\begin{abstract}

We examine the selection characteristics of infrared and sub-mm surveys with \textit{IRAS}, \textit{Spitzer}, BLAST, \textit{Herschel} and SCUBA and identify the range of dust temperatures these surveys are sensitive to, for galaxies in the ULIRG luminosity range (12$<$\,log\,(L$_{\rm IR}$/L$_{\odot}$)\,$<$13), between z=0 and z=4. We find that the extent of the redshift range over which surveys are unbiased is a function of the wavelength of selection, flux density limit and ULIRG luminosity. Short wavelength ($\lambda$\,$\lesssim$200\,$\mu$m) surveys with \textit{IRAS}, \textit{Spitzer}/MIPS and \textit{Herschel}/PACS are sensitive to all SED types in a large temperature interval (17-87\,K), over a substantial fraction of their accessible redshift range. On the other hand, long wavelength ($\lambda$\,$\gtrsim$200\,$\mu$m) surveys with BLAST, \textit{Herschel}/SPIRE and SCUBA are significantly more sensitive to cold ULIRGs, disfavouring warmer SEDs even at low redshifts. In order to evaluate observations in the context of survey selection effects, we examine the temperature distribution of 4 ULIRG samples from \textit{IRAS}, \textit{Spitzer}/MIPS, BLAST and SCUBA. We find that the lack of cold ULIRGs in the local (z$\le$0.1) Universe is not a consequence of the selection and that the range of ULIRG temperatures seen locally is only a subset of a much larger range which exists at high redshift. We demonstrate that the local luminosity-temperature (L-T) relation, which indicates that more luminous sources are also hotter, is not applicable in the distant Universe when extrapolated to the ULIRG regime, because the scatter in observed temperatures is too large. Finally, we show that the difference between the ULIRG temperature distributions locally and at high redshift is not the result of galaxies becoming colder due to an L-T relation which evolves as a function of redshift. Instead, they are consistent with a picture where the evolution of the infrared luminosity function is temperature dependent, i.e. cold galaxies evolve at a faster rate than their warm counterparts. 

\end{abstract}


\section{Introduction}
\label{sec:introduction}

Infrared (IR) radiation was first associated with individual galaxies in the late 1960s (e.g. Johnson 1966\nocite{Johnson66}; Low $\&$ Tucker
1968\nocite{LT68}; Kleinmann $\&$ Low 1970\nocite{KL70}), laying out the path for subsequent observations by the first infrared
space observatory, the InfraRed Astronomical Satellite (\emph{IRAS}). The \textit{IRAS} all sky survey (Soifer, Neugebauer $\&$ Houck 1987)\nocite{SNH87} established the
existence of a class of galaxies whose bolometric energy output emerges almost entirely in the infrared (Soifer et al. 1984a\nocite{Soifer84a}; Sanders $\&$ Mirabel 1996\nocite{SM96}) and amongst those, a population of ultraluminous infrared galaxies (ULIRGs; Soifer et al. 1986; 1987\nocite{Soifer86}\nocite{Soifer87}) with total infrared luminosities (L$_{\rm IR}$, 8--1000\,$\mu$m) greater than 10$^{12}$\,L$_{\odot}$. 

ULIRGs are amongst the most actively star-forming galaxies in the Universe, with a non-trivial fraction of them also hosting AGN ($\gtrsim$\,30 per cent, e.g. Franceschini et al. 2003\nocite{Franceschini03b}, Kartaltepe et al. 2010\nocite{Kartaltepe10}, Symeonidis et al. 2010\nocite{Symeonidis10a}). Studies have shown that although ULIRGs are rare in the local Universe (e.g. Kim $\&$ Sanders 1998\nocite{KS98}), they become more common at high redshift, concomitant with the evolution of the infrared luminosity function and the increase in the total infrared luminosity density with redshift (e.g. Takeuchi, Buat $\&$ Burgarella 2005\nocite{TBB05}). Sub-mm and radio surveys show that ULIRGs are already in place at z$\sim$2--4 (e.g. Kov\'acs et al. 2006\nocite{Kovacs06}; Coppin et al. 2008\nocite{Coppin08}; Seymour et al. 2008\nocite{Seymour08a}) and that together with the luminous infrared galaxy (LIRG, L$_{\rm IR}$=10$^{11}$--10$^{12}$\,L$_{\odot}$) population, they dominate the cosmic infrared luminosity density at z$>$1 (e.g. Lagache et al. 2004\nocite{Lagache04}; P\'erez-Gonz\'alez et al. 2005\nocite{PerezGonzalez05}).

The properties of dust in ULIRGs has been the subject of numerous investigations because, apart from being responsible for most of the energetic output, dust is both a by-product and catalyst for star-formation. The temperature dust can reach in different parts of the interstellar medium (ISM) is related to its distribution, abundance and type, as well as the star-formation rate and efficiency. 
Initial studies showed that ULIRGs were characterised by warm average dust temperatures (e.g. Soifer et al. 1984\nocite{Soifer84b}; Klaas et al. 1997\nocite{Klaas97}). However, these results were based on local ULIRGs only and early observations with the Submillimeter Common User Bolometer Array (SCUBA; Holland et al. 1999) revealed a different picture of the distant Universe (e.g. Hughes et al. 1998\nocite{Hughes98}; Eales et al. 1999\nocite{Eales99}, 2000\nocite{Eales00}). One of the most important outcomes of SCUBA blank-field surveys was discovering the existence of ULIRGs characterised by colder average dust temperatures than those seen locally (Kov\'acs et al. 2006; Pope et al. 2006\nocite{Pope06}; Huynh et al. 2007a\nocite{Huynh07a}; Coppin et al. 2008). 

The abundance of cold ULIRGs at high redshifts has been commonly ascribed to the predisposition of SCUBA surveys to detect such objects. Analogously, our view of local ULIRGs has been influenced by two factors, one of which relates to the potential selection effects introduced by \textit{IRAS} surveys and the other to the apparent correlation between infrared luminosity and dust temperature. Analysis of local spectral energy distributions (SEDs) has revealed that more luminous galaxies are warmer than their lower luminosity counterparts, pointing to a luminosity-temperature (L-T) relation (e.g. Dunne et al. 2000\nocite{Dunne00}; Dale et al. 2001\nocite{Dale01}; Dale $\&$ Helou 2002\nocite{DH02}; Chapman et al. 2003\nocite{Chapman03}), which various works have incorporated in their analysis of the local luminosity function (e.g. Chapman et al. 2003; Blain, Barnard $\&$ Chapman 2003; Lewis, Chapman $\&$ Helou 2005\nocite{LCH05}; Chapin, Hughes $\&$ Aretxaga 2009). At the same time however, it has been suggested that the lack of cold ULIRGs in the local Universe could be due to a selection effect, whereby the \textit{IRAS} 60$\mu$m waveband is expected to favour the selection of sources with warmer dust temperatures (e.g. Blain et al. 2004a\nocite{Blain04a}). 

Recently some progress in resolving these issues has been made. Clements, Dunne $\&$ Eales (2010\nocite{CDE10}) demonstrated that the submm properties of local ULIRGs are consistent with previous results which described them as having warmer temperatures than sub-mm galaxies (SMGs) of comparable luminosity. In addition, Symeonidis et al. (2009\nocite{Symeonidis09}) and Seymour et al. (2010\nocite{Seymour10}) showed that cold ULIRGs are not unique to the redshifts probed by SCUBA; they are already present at z$\lesssim$\,1 and seem to be the main contributors in the strong evolution of the ULIRG energy density with redshift. Nevertheless, until now, the variation in the sensitivity of different instruments has meant that ULIRG samples have not been uniformly mapped out in redshift. \emph{Herschel} (Pilbratt et al. 2010\nocite{Pilbratt10}) is expected to enable a more quantitative comparison than previously possible between local and distant ULIRGs, but such an endeavour will require a fundamental understanding of a) the range of ULIRG SED types and b) what slice of the ULIRG population each survey can probe. 

In this paper we address these topics by firstly focusing on the \textit{IRAS} view of the local Universe (Section \ref{sec:ULIRGs_IRAS}): the temperature range that \textit{IRAS} surveys are sensitive to and the SED types of local ULIRGs (characterised by their peak wavelength, $\lambda_{\rm peak}$). We subsequently examine what types of ULIRG SEDs are traced by \textit{Spitzer}, BLAST and SCUBA at z\,$\le$\,4 and investigate the differences between the local and high redshift samples (Section \ref{sec:highz}). In Section \ref{sec:selection_future}, we examine the potential of \textit{Herschel} and SCUBA\,2 surveys to characterise the ULIRG population in the distant Universe and finally, our discussion and conclusions are presented in Sections \ref{sec:discussion} and \ref{sec:conclusions}. Note that in order to discuss the differences in $\lambda_{\rm peak}$ distributions between the samples we examine in this paper, we occasionally adopt a definition of \textit{warm} and \textit{cold} SEDs as those that peak shortwards and longwards of 90\,$\mu$m respectively. Although, this division is not used as a substitute for a more quantifiable characterisation of the populations, it allows us to place our results in the context of earlier work, where the warm and cold terminology (or variations of it, such as``cool'' or ``hot'') is used extensively to qualitatively describe infrared SEDs (e.g. Dunne et al. 2000\nocite{Dunne00}; Chapman et al. 2002\nocite{Chapman02}, 2003\nocite{Chapman03}; Xilouris et al. 2004\nocite{Xilouris04}; Casey et al. 2009\nocite{Casey09}). Our choice of 90\,$\mu$m corresponds to a greybody temperature of 29\,K which is (i) roughly compatible with the transition wavelength between IR emission from warm dust in young star-forming regions and a cooler component associated with more extended dust heated by the interstellar radiation field (Lonsdale-Persson $\&$ Helou 1987\nocite{LPH87}; see also Seymour et al. 2010); (ii) consistent with previous studies of local ULIRGs where the cold dust component is placed below 30\,K (Klaas et al. 2001\nocite{Klaas01}, Spinoglio, Andreani $\&$ Malkan 2002\nocite{SAM02}) and 28\,K (Hawkins et al. 2001\nocite{Hawkins01}) and the warm component above 30\,K (Vlahakis, Dunne $\&$ Eale 2005\nocite{VDE05}); (iii) in the middle of the range of temperatures probed by``cold'' high-z SCUBA SMGs and ``warm'' local (\textit{IRAS}) ULIRGs (e.g. Clements, Dunne $\&$ Eales 2010\nocite{CDE10}) and (iv) about halfway in the range of SEDs of the Symeonidis et al. (2009) 70\,$\mu$m sample. 

Throughout we employ a concordance consmology of H$_0$=70\,km\,s$^{-1}$Mpc$^{-1}$, $\Omega_{\rm M}$=1-$\Omega_{\rm \Lambda}$=0.3.

\begin{figure*}
\epsfig{file=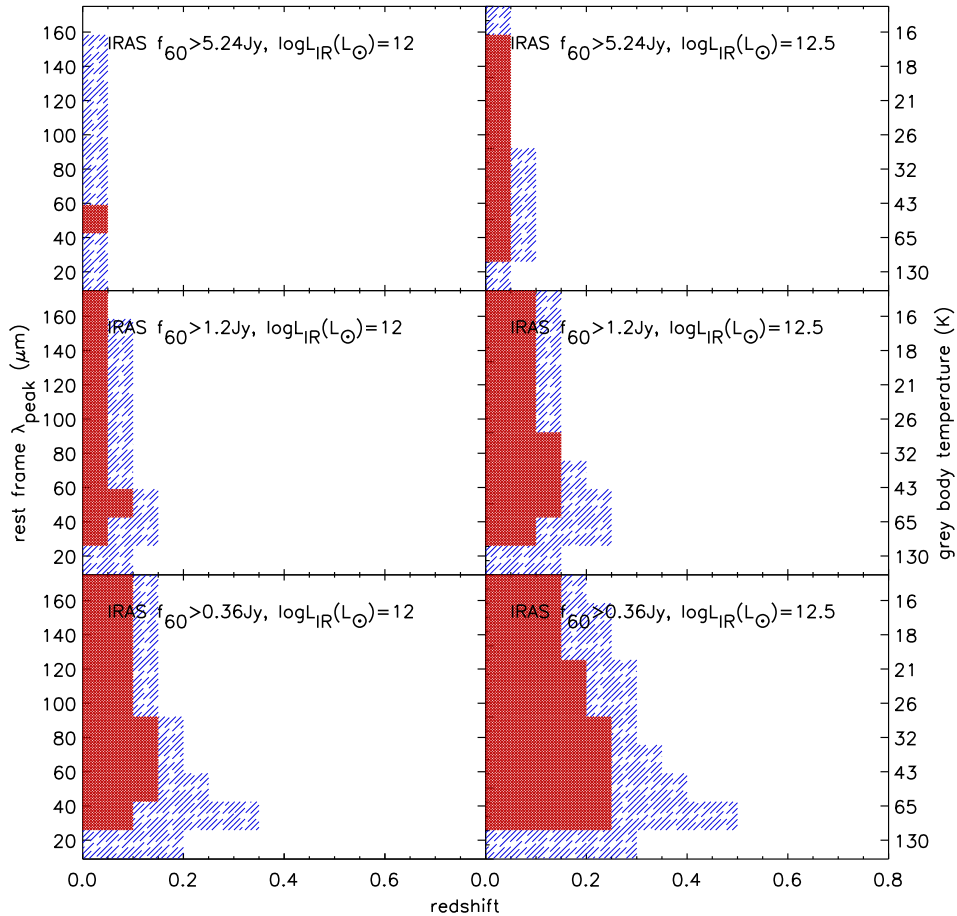,width=13cm}
\caption{Plots of SED peak wavelength ($\lambda_{\rm peak}$) (and temperature, right-hand y-axis)  vs redshift for surveys with \textit{IRAS} for three 60\,$\mu$m flux density limits, 5.24, 1.2 and 0.36\,Jy, corresponding to the BGS, the Fisher et al. (1995) sample and the IIFSCz sample respectively. Left panels and right panels correspond to log\,L$_{\rm IR}$=12 and log\,L$_{\rm IR}$=12.5 ULIRGs respectively. Thick red patterns designate the region where all templates with a certain $\lambda_{\rm peak}$ are detected at a given redshift, whereas unshaded areas correspond regions where no templates are detected. Blue dashed regions indicate that for a given $\lambda_{\rm peak}$ only a fraction of the SK07 templates are recovered. }
\label{fig:fig1}
\end{figure*}

\section{Local ULIRGs as selected by \textit{IRAS}}
\label{sec:ULIRGs_IRAS}

In Symeonidis et al. (2009, hereafter S09) we examined the SEDs of local (z\,$\le$\,0.1) galaxies selected at 60\,$\mu$m down to 5.24\,Jy from the \textit{IRAS}  Bright Galaxy Sample (BGS, Sanders et al. 2003\nocite{Sanders03}). Their rest frame, [$\nu$\,L$_{\nu}$] SED peak wavelength ($\lambda_{\rm peak}$) and total infrared luminosity (L$_{\rm IR}$, 8--1000\,$\mu$m) were estimated by fitting available photometry with the Siebenmorgen $\&$ Kr{\"u}gel (2007, hereafter SK07\nocite{SK07}) SED templates, adopting $\lambda_{\rm peak}$ as a proxy for dust temperature. We found that for a given infrared luminosity, local ULIRGs showed warmer SEDs which peak at shorter wavelengths than what was observed for a sample of higher redshift (0.1$<$z$<$1.2) \textit{Spitzer}/MIPS 70\,$\mu$m-selected sources. Here we re-examine selection issues and opt to provide a more stringent analysis of local galaxy SEDs (i) by investigating the selection characteristics of \textit{IRAS} surveys and (ii) by extending our work to much fainter \textit{IRAS} fluxes. 

We examine the sensitivity of 60\,$\mu$m \textit{IRAS} surveys to various SED types as a function of redshift, for a given flux density limit. SEDs are characterised by their peak wavelength, however to facilitate interpretation of our results, we relate $\lambda_{\rm peak}$ to temperature, using the Wien displacement law for a $\nu$f$_{\nu}$ grey body of dust emissivity $\beta$=1.5:
\begin{equation}
T (K) \sim \frac{hc}{(4+\beta) \times k \times \lambda_{\rm peak}}
\label{wien}
\end{equation}
where h is the Planck constant, c is the speed of light in a vacuum and k is the Boltzmann constant.
We choose 60$\mu$m flux density limits of 5.24, 1.2 and 0.36\,Jy, corresponding to the BGS, the Fisher et al. (1995\nocite{Fisher95}) sample and the Imperial \emph{IRAS}-FSC Redshift Catalogue (IIFSCz; see below) respectively. For our purposes, we use the entire SK07 library, composed of 7088 SED templates which peak in the 9.4--175\,$\mu$m range. Many templates are characterised by more than one dust temperature component and hence the SK07 library is able to match a large variety of systems. We characterise the SEDs via their peak wavelength which is a unique feature and represents the region where the bulk of the emission originates. Figure \ref{fig:fig1} is created as follows: (i) all templates are normalised to log\,(L$_{\rm IR}$/L$_{\odot}$, 8--1000\,$\mu$m)=12 and 12.5, (ii) they are redshifted to z=0.05 up to z=0.8 in steps of 0.05 and (iii) the observed 60\,$\mu$m flux density is estimated for all templates at all redshifts. The $\lambda_{\rm peak}$-z parameter space in figure \ref{fig:fig1} illustrates the fraction of templates with a certain $\lambda_{\rm peak}$ that are detected at a given redshift. Red thick patterns mark the regions where all templates are detected, whereas the blue dashed pattern indicates intermediate regions where only a fraction of the SK07 templates are recovered. In terms of the latter, the detection rate relates to variations in SED shape: when 60\,$\mu$m probes the Wien side of the SED, only SEDs with a shallow mid-IR slope are likely to be detected close to the surveys' flux density limit, whereas when it probes the Rayleigh-Jeans side, only templates with a substantial amount of flux after the peak will be recovered. In the unshaded areas, no templates are detected for a given flux density limit and ULIRG luminosity. 
For our purposes, we classify a survey as unbiased and complete if it can detect all SEDs between 30--150\,$\mu$m (17--87\,K). 

Figure \ref{fig:fig1} demonstrates that the two deeper IRAS surveys are unbiased over certain redshift ranges. Depending on the ULIRG luminosity, this extends to z=0.05 and z=0.1 for the 1.2\,Jy survey and to z=0.1 and z=0.15 for the 0.36\,Jy survey, although for the latter, ULIRGs with SED peak wavelengths in the 30--130\,$\mu$m range can be detected up to z=0.2. For the shallower BGS, the detection of log\,(L$_{\rm IR}$/L$_{\odot}$)$\sim$12 ULIRGs is only complete at hotter temperatures ($\lambda_{\rm peak}$\,=40--60\,$\mu$m), but the same survey is unbiased up to z=0.05 for sources in the more luminous, log\,(L$_{\rm IR}$/L$_{\odot}$)$\sim$12.5, range. 

In the context of the selection function in Fig. \ref{fig:fig1}, we examine the SEDs of \textit{IRAS}-selected ULIRGs, using the Imperial \emph{IRAS}-FSC Redshift Catalogue (IIFSCz, Wang $\&$ Rowan-Robinson 2009\nocite{WRR09}), which is a sub-sample of the \emph{IRAS} Faint Source Catalogue (FSC, Moshir 1992\nocite{Moshir92}) and about 14 times deeper than the BGS. We keep all sources down to a 60\,$\mu$m flux density limit of 0.36\,Jy, where the sample is 90 per cent complete with 90 per cent of the sources having either spectroscopic or photometric redshifts. For SED analysis, we also require 25 and 100\,$\mu$m detections and calculate L$_{\rm IR}$ and $\lambda_{\rm peak}$ by fitting the SK07 models as outlined in S09, after colour-correcting the photometry as specified in the \textit{IRAS} Explanatory Supplement (Beichman et al. 1988\nocite{Beichman88}). The effects of the additional 25 and 100\,$\mu$m selection are assessed as follows: using the 60\,$\mu$m flux density we calculate L$_{\rm IR}$ according to L$_{\rm IR}$=1.7$\times$\,$\lambda$L$_{60}$ (Chapman et al. 2000\nocite{Chapman00}) and examine the 25 and 100\,$\mu$m detection rate of sources in the ULIRG (12$<$\,log\,(L$_{\rm IR}$/L$_{\odot}$)\,$<$13) luminosity range. We find that z$<$0.1 ULIRGs are detected in all 3 bands, however for z$>$0.1 sources the detection rate is 75 per cent at 100\,$\mu$m, with 25 per cent of those also having 25\,$\mu$m data. As a result and in order to work with a complete sample, we focus all subsequent analysis on local (z$\le$0.1) ULIRGs only, leaving us with 27 sources --- cf. 15 local ULIRGs in the BGS in S09.  

\begin{figure}
\epsfig{file=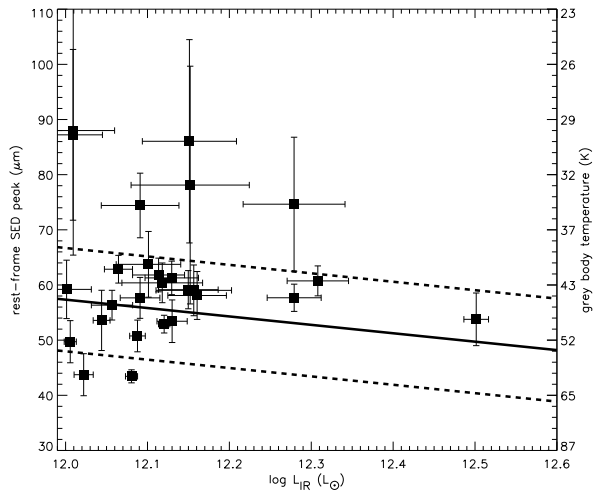,width=8.5cm}
\caption{SED peak wavelength (and dust temperature derived using equation \ref{wien} on the right-hand y-axis) versus total infrared
 luminosity (8--1000\,$\mu$m) for the 27 local (z$\le$0.1) ULIRGs selected at 60\,$\mu$m (f$_{60}$\,$>$\,0.36\,Jy) from the Imperial \emph{IRAS}-FSC Redshift catalogue. The solid and dashed lines are the local luminosity-colour relation and 1$\sigma$ limits from Chapin, Hughes $\&$ Aretxaga (2009), converted into a luminosity-$\lambda_{\rm peak}$ relation using equation \ref{appendix_equation} in Appendix \ref{appendixA}. Note that cold ULIRGs are rare in the local Universe. }
\label{fig:fig2}
\end{figure}

Figure \ref{fig:fig2} shows the rest-frame SED peak wavelength (grey body temperature on the right-hand y-axis) against total infrared luminosity (L$_{\rm IR}$, 8--1000\,$\mu$m) for the 27 local IIFSCz ULIRGs. For comparison we also include the local luminosity-colour relation from Chapin, Hughes $\&$ Aretxaga (2009\nocite{CHA09}, hereafter CHA09). Colour, defined as log\,[f$_{60}$/f$_{100}$], is equivalent to temperature, so this relation describes sources as being hotter if more luminous; see Appendix \ref{appendixA} for details on converting colour to SED peak wavelength (and temperature).  

The IIFSCz comprises about 50 per cent more ULIRGs at z$<$0.1 than the S09 BGS, including some sources with $\lambda_{\rm peak}$\,$>$\,70\,$\mu$m, however, the majority (80 per cent) are consistent with the temperature distribution we saw in S09. In addition, 70 per cent are within 1$\sigma$ of the local luminosity-temperature (L-T) relation of CHA09.
Despite the fact that the 0.36\,Jy survey is sensitive to all ULIRG SED types at z$\le$0.1 (Fig. \ref{fig:fig1}), the average peak wavelength is short: $\left\langle \lambda_{\rm peak} \right\rangle$\,=62$\pm$2\,$\mu$m with an intrinsic dispersion $\sigma$=12\,$\mu$m; $\left\langle T \right\rangle$\,=42$\pm$2\,K with $\sigma$=8\,K. In addition, $\lambda_{\rm peak}$ extends only up to 90\,$\mu$m (albeit large errors on some sources). Our results confirm that cold ULIRGs are rare in the local Universe (see section \ref{sec:introduction} for our definition of cold).

\begin{figure*}
\epsfig{file=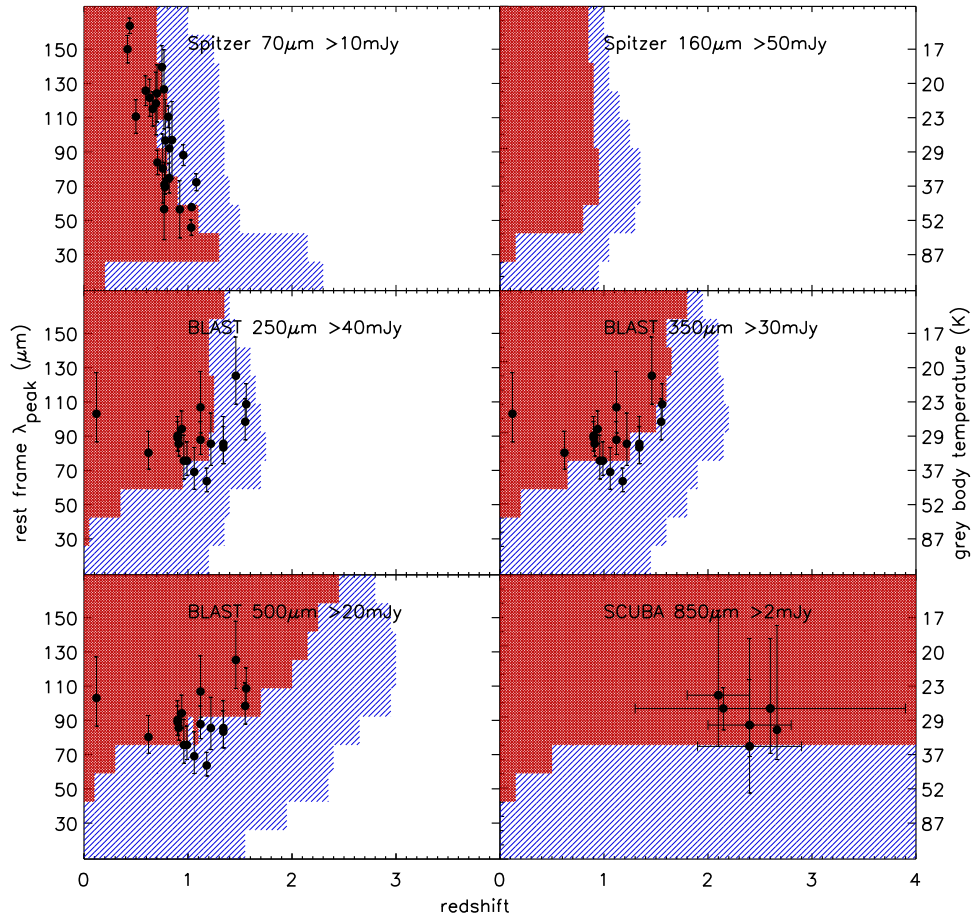,width=13cm}
\caption{Plots of SED peak wavelength (and grey body temperature on the right-hand y-axis) vs redshift for surveys with \textit{Spitzer}, BLAST and SCUBA for log\,(L$_{\rm IR}$/L$_{\odot}$)=12.5 ULIRGs, using flux density limits of 10, 50, 40, 30, 20 and 2\,mJy for the 70, 160, 250, 350, 500 and 850\,$\mu$m bands respectively. We also plot all log\,L$_{\rm IR}$\,$\le$12.5 ULIRGs from the samples of Symeonidis et al. (2009) (70 and 160 \,$\mu$m panels), Dye et al. (2009) (250, 350 and 500\,$\mu$m panels) and Coppin et al. (2008) (850\,$\mu$m panel). In the red thickly patterned regions all templates with a particular $\lambda_{\rm peak}$ are detected at a given redshift, whereas white areas indicate that no templates are detected. Blue dashed regions indicate that only some of the SK07 templates with a particular $\lambda_{\rm peak}$ are detected at that redshift.}
\label{fig:fig3}
\end{figure*}

\begin{figure*}
\epsfig{file=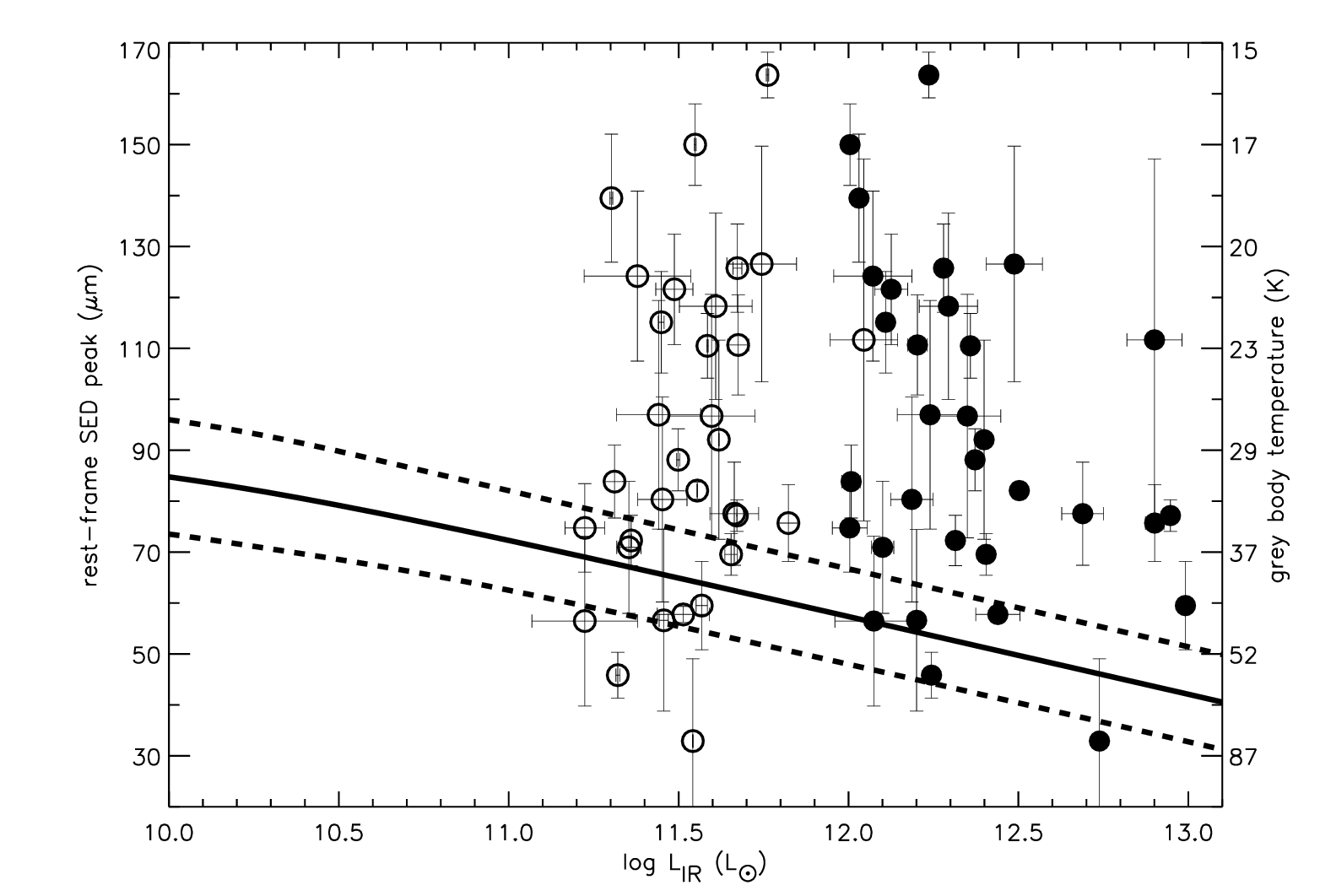,width=12cm} 
\caption{SED peak wavelength (grey body temperature on the right-hand y-axis) versus total infrared luminosity for the \textit{Spitzer}/70$\mu$m sample of Symeonidis et al. (2009), which is at 0.4$<$z$<$1.2, with $\left\langle \rm z \right\rangle$=0.9. The solid and dashed lines are the L-T relation and 1$\sigma$ limits from Chapin, Hughes $\&$ Aretxaga (2009). Filled symbols indicate the true luminosities. Empty symbols correspond to the luminosities divided by (1+z)$^3$. Here, we demostrate that evolution of the L-T relation with redshift cannot be responsible for the increased presence of cold ULIRGs and range of temperatures we see in the distant Universe.}
\label{fig:fig4}
\end{figure*}

\section{ULIRGs at high redshift}
\label{sec:highz}

\subsection{The selection function of \textit{Spitzer}, BLAST and SCUBA}
\label{sec:selection_current}

To quantitatively describe the selection issues relating to various infrared and submm surveys, we develop a schematic representation of the selection functions of Balloon-borne Large Aperture Submm Telescope (BLAST; Pascale et al. 2008\nocite{Pascale08}), SCUBA and \textit{Spitzer} 70 and 160\,$\mu$m surveys for log\,(L$_{\rm IR}$/L$_{\odot}$)\,=\,12.5 ULIRGs up to z=4. Figure \ref{fig:fig3} is built following the method outlined in section \ref{sec:ULIRGs_IRAS}, where the $\lambda_{\rm peak}$-z parameter space illustrates the temperatures these surveys are sensitive to at any given redshift and for a given flux density limit. As in Fig. \ref{fig:fig1}, unshaded regions and those shaded red correspond to no and all SK07 templates being detected respectively. In the blue dashed areas, only some of the templates are recovered. For the submm surveys (BLAST and SCUBA), this corresponds to the detection of templates with significant flux in the wavelength range probed by these instruments, which is always longward of the peak. For \textit{Spitzer} 70 and 160\,$\mu$m the same applies only if the SED peaks shortwards of the wavelength of selection, otherwise it is the mid-IR continuum slope that determines the detection rate (see also section \ref{sec:ULIRGs_IRAS}). 
The flux density limits that we use for each survey are: 10\,mJy for MIPS 70\,$\mu$m, matching our observations in S09, 50\,mJy for MIPS 160\,$\mu$m (Dole et al. 2004\nocite{Dole04}), 40, 30 and 20\,mJy for the 250, 350 and 500\,$\mu$m BLAST bands respectively (Dye et al. 2009\nocite{Dye09}, hereafter D09) and 2\,mJy for SCUBA (Coppin et al. 2008\nocite{Coppin08}, hereafter C08). In Fig. \ref{fig:fig3}, we also overlay data for log\,(L$_{\rm IR}$/L$_{\odot}$)\,$\le$12.5 ULIRGs from the S09 70\,$\mu$m sample, SCUBA measurements from Coppin et al. (2008) and BLAST measurements from Dye et al. (2009). For the SCUBA and BLAST data we use equation \ref{wien} to convert their derived temperatures to $\lambda_{\rm peak}$. Ideally, we should be looking at ULIRGs in a narrow slice around log\,(L$_{\rm IR}$/L$_{\odot}$)\,=12.5 to be consistent with the selection function displayed, however the samples are too small and hence we plot all 12$\le$\,log\,(L$_{\rm IR}$/L$_{\odot}$)$\le$12.5 sources. In addition, all subsequent estimates of the average peak wavelength and temperature refer to these log\,(L$_{\rm IR}$/L$_{\odot}$)$\le$12.5 sources only, in order to achieve a fair comparison with local ULIRGs whose luminosities do not extend beyond log\,(L$_{\rm IR}$/L$_{\odot}$)=12.5.

The MIPS 70\,$\mu$m survey can potentially detect objects up to z=2, but with severe incompleteness above z=1 --- it is unbiased up to z\,$\sim$0.7 and sensitive to warmer SEDs above that redshift. Our predictions are consistent with results from Casey et al. (2009\nocite{Casey09}) who find all z$>$1, 70\,$\mu$m-detected ULIRGs to be hot. Surveys at 160\,$\mu$m have a similar selection function to the ones at 70\,$\mu$m, but with a temperature range which varies less strongly with redshift, apart from the very hot ($\lambda_{\rm peak}$\,$<$\,40\,$\mu$m) SED region. On the other hand for the longer wavelength surveys the likelihood of detection is significantly more dependent on SED type. BLAST is highly sensitive to colder SEDs, but sources with $\lambda_{\rm peak}$=\,60--90\,$\mu$m can still be detected at 250\,$\mu$m up to z=1. Down to 2\,mJy SCUBA should be able to detect ULIRGs over the entire $\lambda_{\rm peak}$-z parameter space, primarily because of the negative \textit{K}-correction which offers a large advantage above z\,$\sim$\,1. However above z=0.5, there is a strong predisposition towards colder ($\lambda_{\rm peak}$$>$70\,$\mu$m) objects. In addition, the large area below $\lambda_{\rm peak}$=70\,$\mu$m covered by the blue regions in the BLAST and SCUBA panels suggests that warm SEDs can only be recovered if they have sufficient submm flux, i.e. an additional cold temperature component. 

How does the dust temperature distribution of the BLAST, SCUBA and \textit{Spitzer} samples compare in the context of their selection characteristics? The S09 sample's $\lambda_{\rm peak}$ distribution shifts from longer to shorter values with increasing redshift, following the shape of the selection function, which shows increased sensitivity towards warm SEDs above z=0.7. Although above z=0.8, where the survey becomes more sensitive to $\lambda_{\rm peak}$\,$<$\,70\,$\mu$m, most ULIRGs are warm, it is interesting to note that in the unbiased region (z$<$0.7) most ULIRGs are cold. We calculate $\left\langle \lambda_{\rm peak} \right\rangle$\,=99$\pm$6\,$\mu$m with an intrinsic dispersion $\sigma$=31\,$\mu$m; $\left\langle T \right\rangle$\,=26$\pm$2\,K with $\sigma$=8\,K. 

The D09 sample consists of sources detected at 5$\sigma$ in at least one BLAST band. Almost all are also detected at 70 and 160\,$\mu$m, although the MIPS counterparts have fainter flux densities than the S09 sample. The average peak wavelength of the D09 sample is consistent with that of S09, although the characteristics of the BLAST ULIRGs cannot be evaluated in an unbiased part of the $\lambda_{\rm peak}$-z space. For D09 we calculate $\left\langle \lambda_{\rm peak} \right\rangle$\,=89$\pm$4 with an intrinsic dispersion $\sigma$=15\,$\mu$m; $\left\langle T \right\rangle$\,=29$\pm$1\,K with $\sigma$=5\,K. Note that the range in $\lambda_{\rm peak}$ is slightly different for the MIPS and BLAST samples, with the former extending to lower and higher $\lambda_{\rm peak}$. In addition, their redshift distributions are different, with most S09 ULIRGs being below z=1 and most D09 ULIRGs above z=1. 
The average peak wavelength of the C08 SCUBA sample is also in agreement with D09 and S09. For C08 we calculate $\left\langle \lambda_{\rm peak} \right\rangle$\,=91$\pm$4\,$\mu$m with an intrinsic dispersion $\sigma$=11\,$\mu$m and $\left\langle T \right\rangle$\,=29$\pm$1\,K with $\sigma$=3\,K, however note that (i) there are large upper errors on the temperature measurements, (ii) we are not probing an unbiased part in the selection function and (iii) there are additional selection effects due to the radio selection (see also Blain et al. 2004b\nocite{Blain04b}; Coppin et al. 2009\nocite{Coppin09}). 
We point out that the $\lambda_{\rm peak}$ distribution of these high redshift samples extends to much longer peak wavelengths than what is seen for the local ULIRGs in Fig. \ref{fig:fig2}, resulting in a higher average $\lambda_{\rm peak}$ by at least 2\,$\sigma$ (here $\sigma$ is the dispersion in the local sample, equal to 12\,$\mu$m).

\begin{figure*}
\centering
\begin{tabular}{c|c}
\epsfig{file=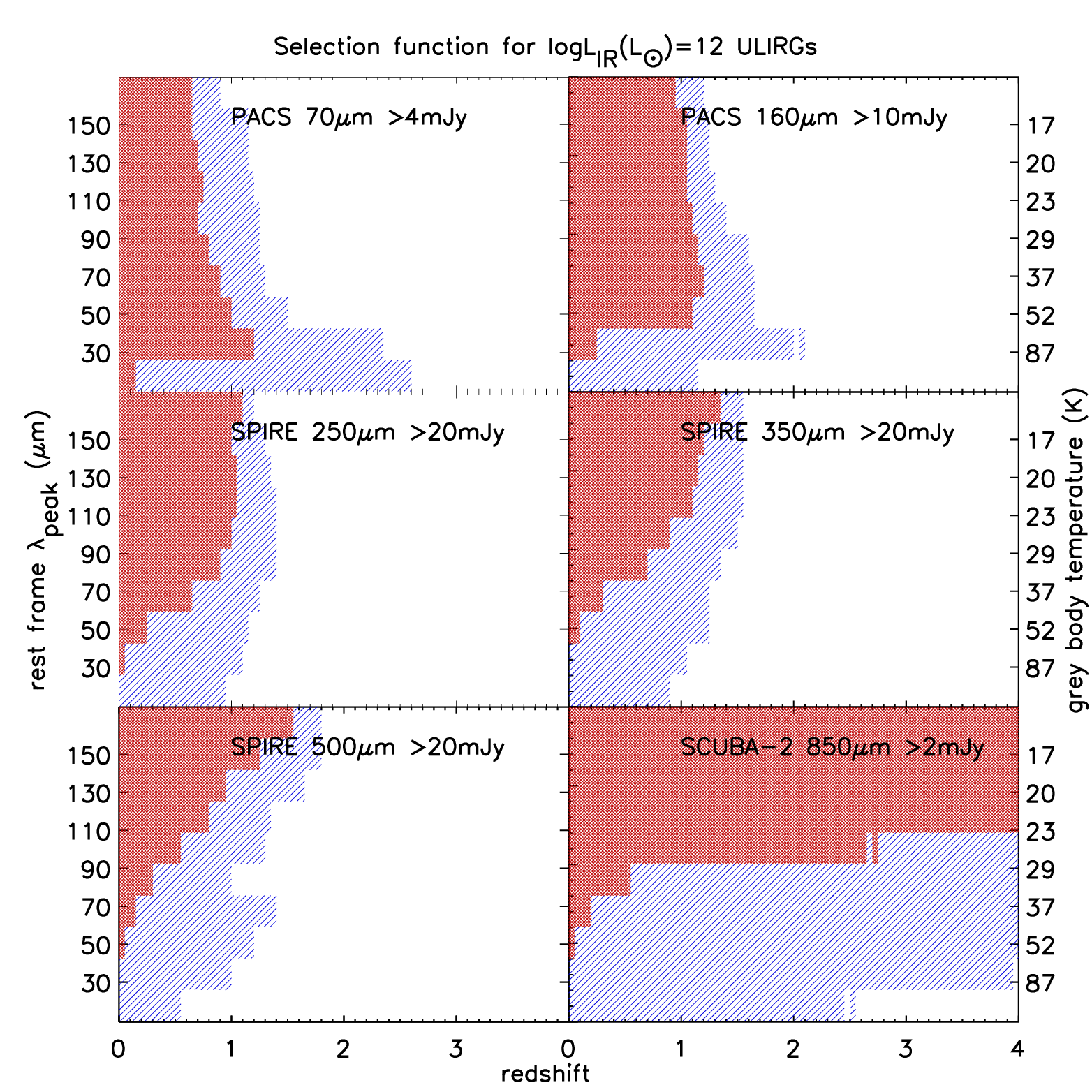,width=0.5\linewidth,clip=} & \epsfig{file=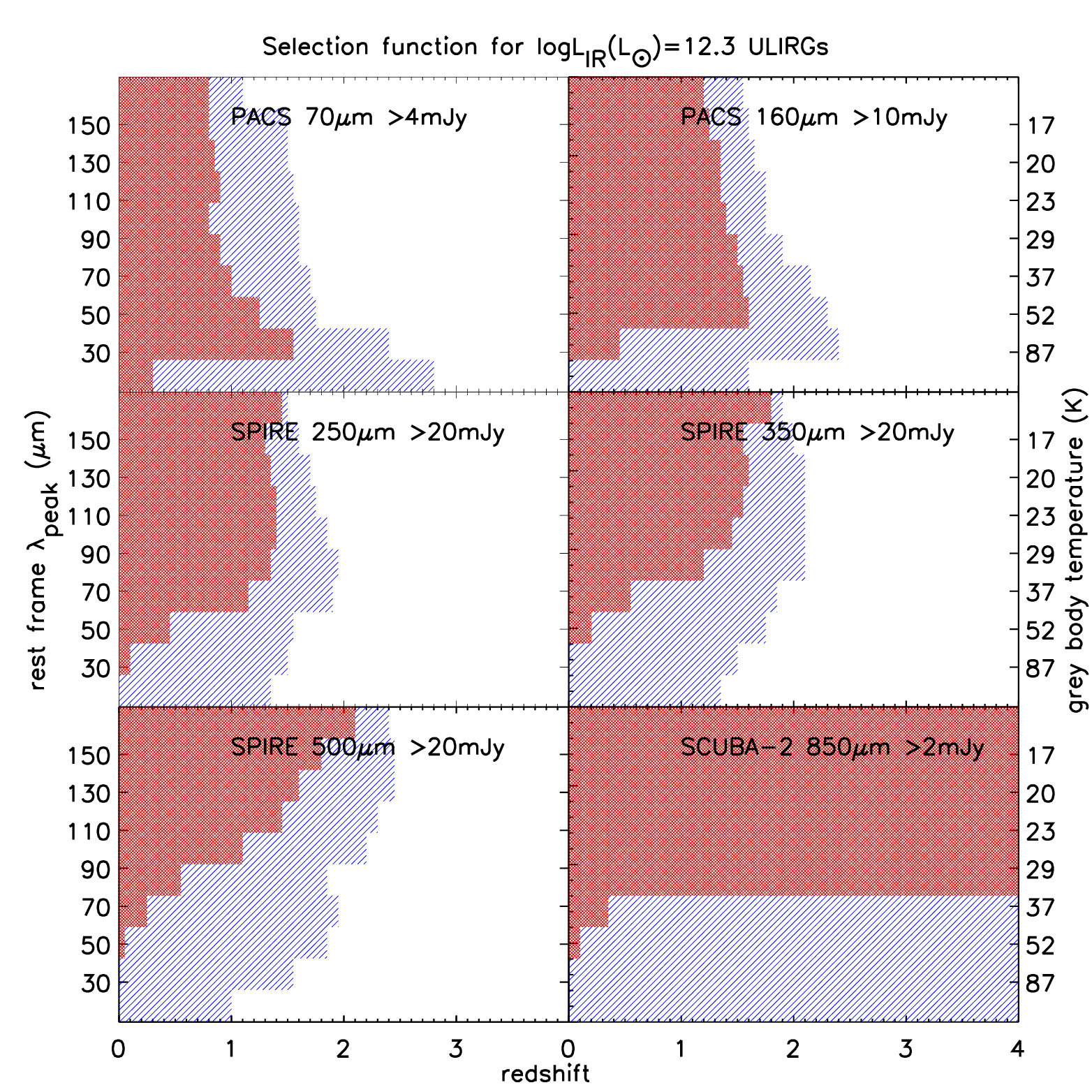,width=0.5\linewidth,clip=} \\
\epsfig{file=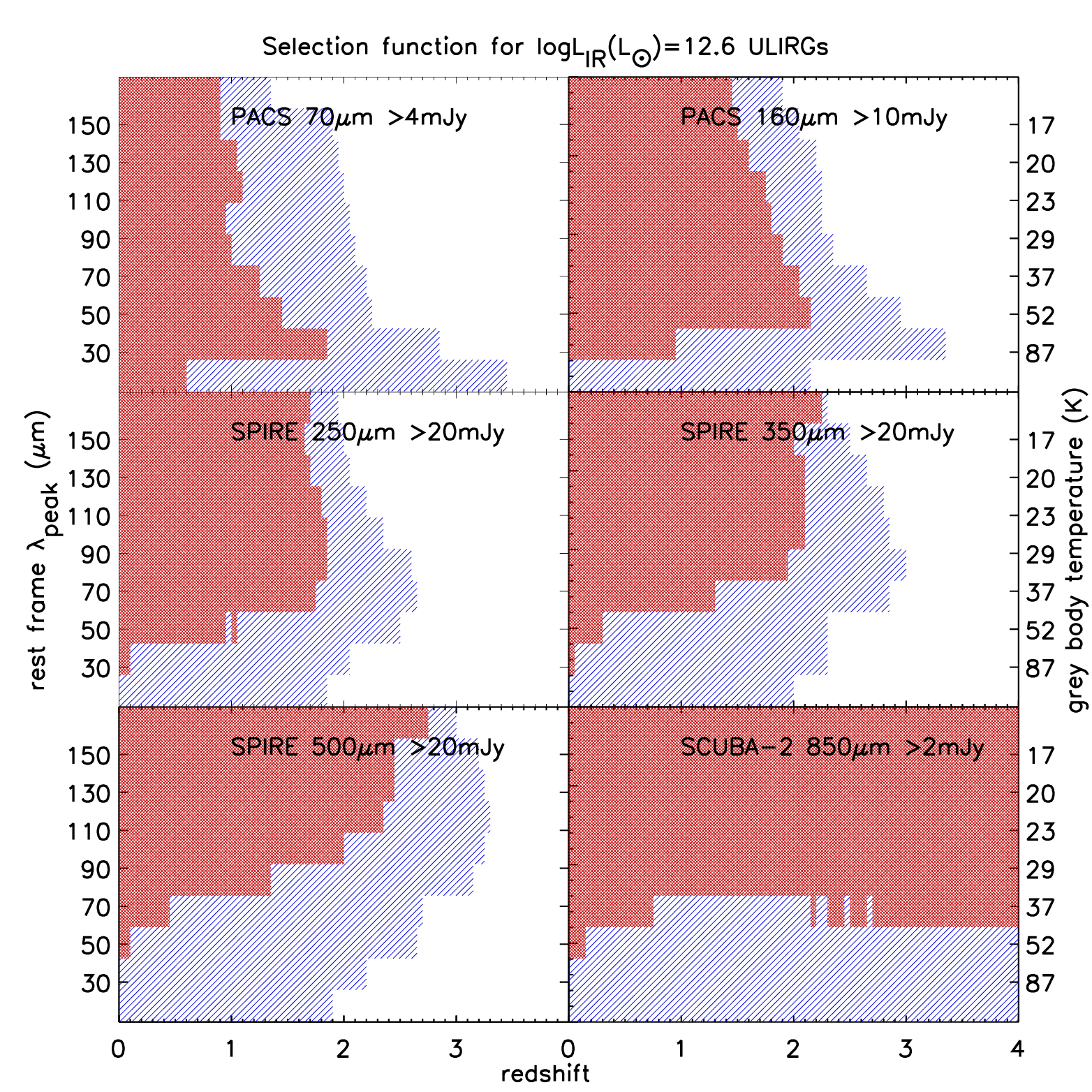,width=0.5\linewidth,clip=} &\epsfig{file=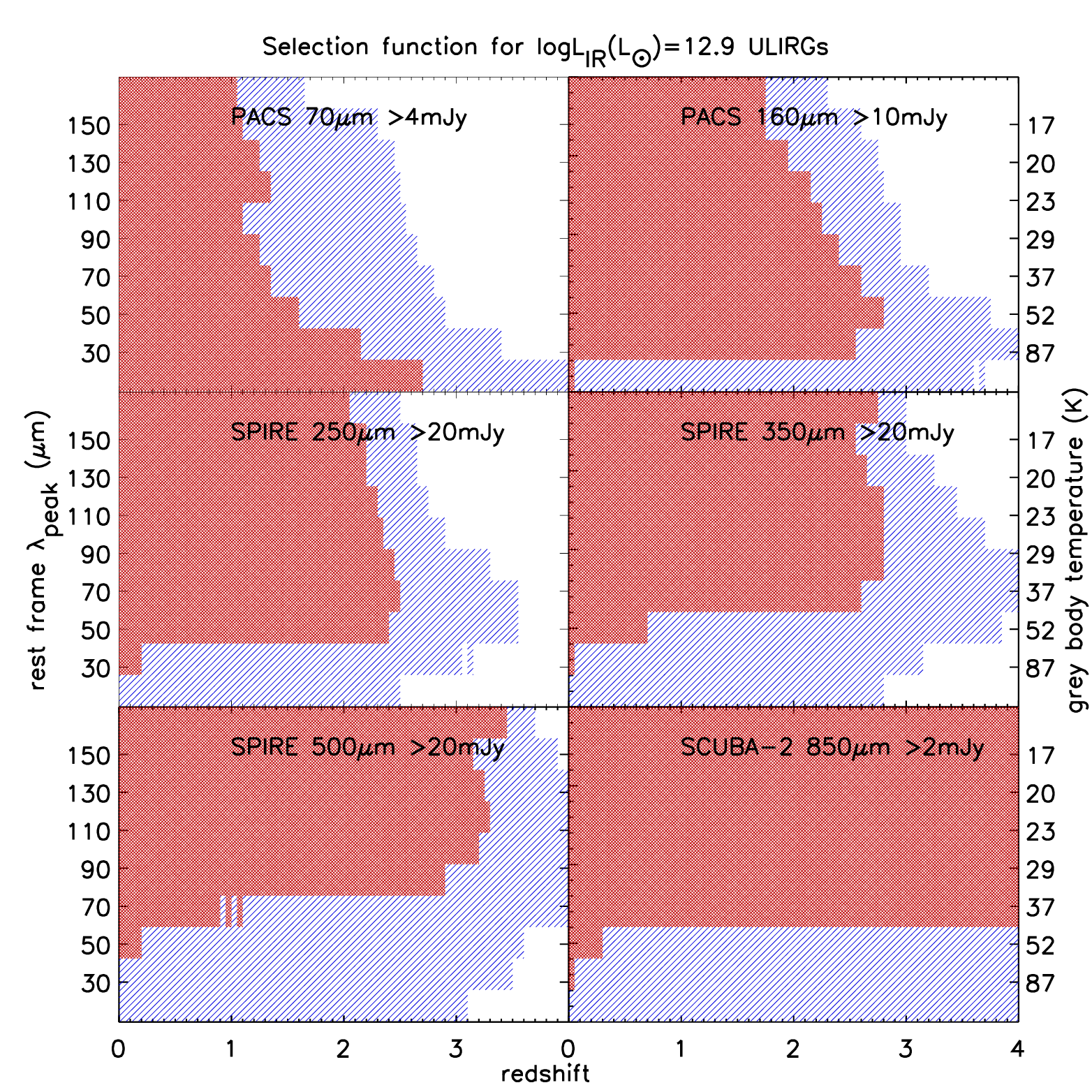,width=0.5\linewidth,clip=} \\
\end{tabular}
\caption{The $\lambda_{\rm peak}$-redshift parameter space for surveys with \textit{Herschel} and SCUBA for log\,(L$_{\rm IR}$/L$_{\odot}$)\,=\,12, 12.3, 12.6 and 12.9 ULIRGs. For more details, see caption in figures \ref{fig:fig1} and \ref{fig:fig3}.}
\label{fig:fig5678}
\end{figure*}

\subsection{An L-T relation at high redshift?}
\label{sec:LT_highz}

In S09 we showed that a luminosity-temperature correlation is not apparent in the 70$\mu$m sample, although it is clearly seen in the \textit{IRAS} BGS. However, it has been proposed that the abundance of cold galaxies in the distant Universe is a consequence of the local L-T relation evolving with redshift (e.g. Chapman et al. 2002\nocite{Chapman02}; Lewis, Chapman and Helou 2005\nocite{LCH05}; CHA09) in the same way that the knee of the infrared luminosity function evolves by (1+z)$^q$, where q$\sim$3 (e.g. Le Floc'h et al. 2005\nocite{LeFloch05}, Huynh et al. 2007b\nocite{Huynh07b}). This suggests that although more luminous galaxies would still have higher temperatures than their less luminous counterparts, they would be much colder than the objects we see locally, i.e. the ULIRG temperature distribution would shift towards longer $\lambda_{\rm peak}$ (lower T) with increasing redshift. 

In Fig. \ref{fig:fig4}, we examine the applicability of this hypothesis by de-evolving the luminosities of all ULIRGs (12$<$\,log\,(L$_{\rm IR}$/L$_{\odot}$)\,$<$13) in the S09 sample by (1+z)$^3$. CHA09 find that once the luminosities of SCUBA sources are de-evolved they become consistent with the local L-T relation. The same is not true for our lower redshift S09 ULIRGs (0.4$<$z$<$1.2 and $\left\langle \rm z \right\rangle$=0.9), as most remain well outside the relation, which fails to accomodate the large scatter in their temperatures. This result indicates that evolution of the local L-T relation in the ULIRG regime cannot be responsible for the increased presence of cold ULIRGs and the range of SEDs we see in the distant Universe.

\section{Selection of ULIRGs in \textit{Herschel} and SCUBA\,2 surveys}
\label{sec:selection_future}

In a similar fashion to sections \ref{sec:ULIRGs_IRAS} and \ref{sec:selection_current}, we examine the potential of current surveys to measure the variation in ULIRG SED types by predicting the $\lambda_{\rm peak}$-z parameter space that \textit{Herschel} and SCUBA\,2 will cover in different bands for sources of log\,(L$_{\rm IR}$/L$_{\odot}$)\,=\,12, 12.3, 12.6 and 12.9 (Fig. \ref{fig:fig5678}). For the PACS sensitivity limits we take the values from Poglitsch et al. (2010\nocite{Poglitsch10}): $\sim$4 and 10\,mJy for the 70 and 160$\mu$m bands respectively for a 10\arcmin\,$\times$\,15\arcmin scan map with 5$\sigma$/30h integration. We do not include the 100\,$\mu$m band, as the shape of its selection function is similar to that at 70 and 160\,$\mu$m, effectively a very slow transition between the two bands. The SPIRE (Griffin et al. 2010\nocite{Griffin10}) flux density limits of 20\,$\mu$m for all bands, roughly correspond to the 3\,$\sigma$ confusion limit (Nguyen et al. 2010\nocite{Nguyen10}). Finally we assume a SCUBA\,2 flux density limit of 2\,mJy as for SCUBA in Fig. \ref{fig:fig3}. 

Compared to MIPS 70\,$\mu$m ultradeep data in the Great Observatories Origins Deep Survey (GOODS, e.g. Frayer et al. 2006\nocite{Frayer06b}) and other Far-Infrared Deep Extragalactic Legacy (FIDEL, PI: Mark Dickinson, NOAO) survey fields, observations with the PACS 70\,$\mu$m band will mainly benefit from higher spatial resolution, but not from a significant increase in sensitivity; in terms of the latter, the superiority of PACS over MIPS lies largely at 100 and 160\,$\mu$m. Fig. \ref{fig:fig5678} shows that up to z=1 all ULIRGs will be detected in at least one PACS band. On the other hand, up to the same redshift, SPIRE heavily favours colder SEDs, and ULIRGs with $\lambda_{\rm peak}$$>$90$\mu$m are expected to have counterparts above the 3\,$\sigma$ confusion limit in all bands. With respect to warmer SEDs, the extent of the blue shaded area suggests that they will only be detected if they have enough flux at submm wavelengths, a trend which persists even for the more luminous (log\,L$_{\rm IR}$\,$>$\,12.5) sources, especially at the longer wavelength bands. 

Between 1$<$z$<$2, the PACS 70\,$\mu$m selection becomes heavily dependent on SED shape, with many of the log\,L$_{\rm IR}$\,$<$\,12.5 objects not having 70\,$\mu$m counterparts and a stronger tendency towards warmer SEDs for the more luminous (log\,L$_{\rm IR}$\,$>$\,12.5) ULIRGs. The shape of the 160\,$\mu$m selection function behaves in a similar fashion at low luminosities, however for the more luminous ULIRGs it becomes relatively unbiased with respect to temperature. In the same redshift interval, SPIRE misses many of the lower luminosity sources and displays a strong tendency towards cold SEDs for the more luminous ULIRGs. As a result, the log\,L$_{\rm IR}$\,$>$12.5 sources should be detected in more than one SPIRE band if cold and in more than one PACS band if warm, although with certain SED shapes it is also likely that they will have both PACS and SPIRE counterparts. 

The picture is dramatically different at 2$<$z$<$3, where only the most luminous ULIRGs (log\,L$\sim$12.9, borderline HyLIRGs) will be identified in sufficient numbers. However their detection rate is again dependent on SED shape. For certain SED shapes, e.g. for sources which peak below 90\,$\mu$m but have strong submm flux and a shallow mid-IR continuum, we can expect detections in all \textit{Herschel} bands. Nevertheless, for a large fraction of ULIRGs in that redshift interval there will not be uniform and systematic information available on their SEDs. 

SCUBA\,2 is expected to perform better than its predecessor and due to its much higher survey speed and wider field of view it should be able to fill out the $\lambda_{\rm peak}$-redshift parameter space considerably better. Note that its selection function favours the cold ULIRG population uniformly at all redshifts, however its sensitivity towards warm ULIRGs is minimal, with a strong dependence on SED shape. The warm $\lambda_{\rm peak}$-z parameter space is covered better at high L$_{\rm IR}$, as the more luminous sources will have enough submm flux even if warm. 

It is interesting to note that at high L$_{\rm IR}$ and around z=2, the SPIRE 250\,$\mu$m selection turns over to display increased sensitivity to warmer galaxies, taking on a shape similar to that of the 160\,$\mu$m selection function. At z\,$\ge$\,2, 250\,$\mu$m corresponds to rest-frame $\le$\,83\,$\mu$m, which means that the selection will henceforth favour warmer galaxies. This effect is only apparent in the more luminous sources, as lower luminosity ULIRGs cannot be detected at those redshifts above the 3\,$\sigma$ confusion. A similar turnover also occurs at 160\,$\mu$m, however the transition is faster in this case, as shorter peak wavelengths (and warmer temperatures) are already attainable by z=1. Note that the turnover that various selection functions undergo at high redshifts (and hence high ULIRG luminosities in a flux limited sample) can fabricate a correlation between luminosity and temperature. Nonetheless, this correlation is not an inherent property of high redshift IR populations, so caution should be taken not to confuse it with the intrinsic L-T relation we see in the local Universe (see earlier sections) --- e.g. see Ivison et al. (2010) where sources in a BLAST 250\,$\mu$m-selected sample need to be both more luminous and hotter in order to be detected at higher redshift.

\section{Discussion}
\label{sec:discussion}

We have investigated how selection effects influence our view of the ULIRG population both locally and at high redshift. Using a library of infrared SED models, we examined the selection characteristics of various infared and submm surveys with \textit{IRAS}, \textit{Spitzer}, BLAST, \textit{Herschel} and SCUBA and were able to predict the range of ULIRG SED types detectable at a particular redshift, given a survey's flux density limit and wavelength of selection. We subsequently evaluated observations of local and distant ULIRGs against these predictions. SEDs were characterised in terms of their rest-frame peak wavelength (equivalent to dust temperature) and total infrared luminosity (in the 8--1000\,$\mu$m range). A survey was then classed as complete and unbiased over a redshift range where ULIRG SEDs with peak wavelengths in the 30-150\,$\mu$m range (T=17--87\,K) have equal likelihood of being detected. 

Although the sensitivity of \textit{IRAS} does not enable high redshift observations, our analysis shows that 60\,$\mu$m surveys can be unbiased over a given redshift interval, the size of which increases with L$_{\rm IR}$ and survey depth. More specifically, down to a 60\,$\mu$m flux density limit of 0.36\,Jy, we find that \textit{IRAS} is sensitive to all ULIRG SED types and luminosities in the local (z$\le$0.1) Universe. Surveys with \textit{Spitzer}/MIPS and \textit{Herschel}/PACS which target distant galaxies, have the advantage of remaining unbiased over a large fraction of the redshift range that they probe; on average they are complete up to z$\sim$1 and their selection function displays little dependency on the flux density limit. On the other hand longer wavelength surveys with BLAST, \textit{Herschel}/SPIRE and SCUBA which work in the sub-mm regime, are significantly more sensitive to colder ULIRGs over the majority of their accessible redshift range. This is partly due to confusion in the submm limiting their depth and partly due to the fact that submm wavelengths probe emission from very cold dust at the steep Rayleigh-Jeans part of the SED. 

Down to the 3\,$\sigma$ confusion limit, the synergy between PACS and SPIRE should enable complete characterisation of ULIRG SEDs up to z$\sim$1. Above z$\sim$1, the two instruments will likely sample different parts of the ULIRG population but with significant overlap. For the more luminous (log\,L$_{\rm IR}$\,$>$12.5) sources, surveys with \textit{Herschel} and SCUBA\,2 collectively have the potential to cover the whole distribution in $\lambda_{\rm peak}$ up to z$\sim$2. At redshifts beyond 2 however, characterisation of the ULIRG population becomes highly incomplete, as many warm ULIRGs will not be detected and many cold ULIRGs will potentially have submm counterparts in fewer than 2 bands so their SEDs will not be measurable. 

In the context of the various surveys' selection functions we made comparisons between local and high redshift ULIRGs. Using a near complete sample from the deep 0.36\,Jy 60\,$\mu$m \textit{IRAS} survey, we firstly examined the local (z$\le$0.1) ULIRG temperature distribution. Our results show that, despite a uniform sensitivity to all ULIRG temperatures, the average peak wavelength of local ULIRGs is short --- $\left\langle \lambda_{\rm peak} \right\rangle$\,=62\,$\mu$m; $\left\langle T \right\rangle$\,=42\,K. In the same luminosity range (12$\le$\,log\,(L$_{\rm IR}$/L$_{\odot}$)$\le$12.5), the MIPS/70\,$\mu$m ULIRGs from Symeonidis et al. (2009) are found to be cold up to the redshift where the survey is unbiased (z=0.7), while the whole population has an average peak wavelength of $\left\langle \lambda_{\rm peak}\right\rangle$\,=99\,$\mu$m ($\left\langle T \right\rangle$\,=26\,K), 3$\sigma$ higher than the local sample. 
Although, we are not able to study BLAST (Dye et al. 2009) and SCUBA (Coppin et al. 2008) samples in unbiased redshift ranges, as these are too small for current flux density limits, we find them to have consistent temperatures to the MIPS/70\,$\mu$m sample --- $\left\langle \lambda_{\rm peak} \right\rangle$\,=89 and 91\,$\mu$m ($\left\langle T \right\rangle$\,=29 and 29\,K) respectively. We note that, besides a higher value in the average SED peak wavelength, both the scatter and the range in $\lambda_{\rm peak}$ is larger at high redshift.

Given the differences in the $\lambda_{\rm peak}$ distribution between local and high-redshift ULIRGs, one might also ask whether the correlation between IR luminosity and dust temperature, which has been observed in the local Universe, is also applicable at high redshift. The local L-T relation dictates that in the ULIRG regime the average temperature should be high: the L-T  relation in Chapin, Hughes $\&$ Aretxaga (2009) extrapolated to ULIRGs predicts temperatures in the 40--65\,K range within 1\,$\sigma$. Although our results regarding the local 0.36\,Jy sample are in agreement with this temperature range, the extrapolation in L-T for the ULIRG regime fails to accomodate both the large scatter in temperature and the low average temperature seen at high redshift: (i) only 16 per cent of the 70\,$\mu$m sample lie within the local L-T relation 1\,$\sigma$ limits, whereas no sources in the SCUBA and BLAST samples lie within the local L-T relation and (ii) the 3 samples display no apparent connection between L$_{\rm IR}$ and $\lambda_{\rm peak}$. Including results presented here and in S09, ULIRGs at z\,$\sim$\,1 and beyond which span a large range in dust temperature have been discovered in many surveys. Cold ULIRGs have been identified with ISO (e.g. Chapman et al. 2002\nocite{Chapman02}), \textit{Spitzer} (e.g. Kartaltepe et al. 2010) and SCUBA (e.g. Chapman et al. 2005\nocite{Chapman05}). Warm ULIRGs, many of which do not have appreciable submm detections, but are detected at shorter infrared wavelengths or other parts of the spectrum altogether (such as radio or optical), have also been shown to exist (e.g. Blain et al. 2004a\nocite{Blain04a}; Chapman et al. 2004\nocite{Chapman04}; Bussmann et al. 2009\nocite{Bussmann09}, Casey et al. 2009\nocite{Casey09}, Younger et al. 2009\nocite{Younger09}). The collective results from these studies establish a large range in ULIRG temperatures, suggesting that they do not conform to a universal luminosity-temperature correlation. Note that although such a correlation is sometimes observed (e.g. Blain, Barnard $\&$ Chapman 2003\nocite{BBC03}; Chapman et al. 2004, 2005\nocite{Chapman04}\nocite{Chapman05}; Ivison et al. 2010\nocite{Ivison10}), it does not have a physical origin. Rather, the surveys' selection functions imply that at above a certain redshift, ULIRGs will need to be both more luminous and hotter in order to be detected. In theory, the selection function of all surveys should turn to favour warm sources at some redshift, however the longer the wavelength of selection, the higher is the redshift that this will occur at. 

In order to explain the abundance of cold ULIRGs at high redshift and the lack of them in the local Universe, it has been suggested that the local L-T relation evolves in the same way as the knee of the luminosity function, by (1+z)$^3$, i.e. galaxies become colder at high redshift. However, the large observed scatter in ULIRG temperatures and the lack of an observable trend between luminosity and temperature amongst distant ULIRGs is a strong argument against this proposal; we demonstrate this in Fig. \ref{fig:fig4} and Section \ref{sec:LT_highz}. So how can one characterise the discrepancy in the temperature distributions of local and high redshift samples? A recent study by Seymour et al. (2010) examined the evolution in infrared luminosity density (IRLD) with redshift as a function of dust temperature, by splitting sources into warm and cold (at $\lambda_{\rm peak}$=90\,$\mu$m) and examining their contribution to the IRLD separately. Seymour et al. showed that cold galaxies undergo stronger evolution and are primarily responsible for the increase in infrared luminosity density with redshift, a result which can describe the differences in the ULIRG temperature distributions we see locally and at high redshift. If cold ULIRGs evolve faster than their warm counterparts, then we would expect their relative numbers to diminish with decreasing redshift. Although, the division of galaxies into two groups, warm and cold, is possibly too simplistic, this picture does highlight a possible direction for future work on this topic. With clean samples from carefully tailored surveys over redshift ranges where we have shown them to be unbiased, \textit{Herschel} should be key in fully characterising ULIRG SEDs and hence investigating the evolution of the IR luminosity function as a function of temperature.

\section{Conclusions}
\label{sec:conclusions}

Our conclusions are summarised below:

\begin{enumerate}
\item All surveys, independent of the wavelength of selection, have the potential of being complete and unbiased with respect to ULIRG SEDs over a certain redshift range, the extent of which increases with survey depth and ULIRG luminosity. As it stands, short wavelength ($\lambda$\,$\lesssim$200\,$\mu$m) surveys with \textit{IRAS}, \textit{Spitzer}/MIPS and \textit{Herschel}/PACS are sensitive to all SED types in a large temperature interval (17-87\,K) over a substantial fraction of their accessible redshift range. On the other hand, long wavelength ($\lambda$\,$\gtrsim$200\,$\mu$m) surveys with BLAST, \textit{Herschel}/SPIRE and SCUBA are significantly more sensitive to cold ULIRGs, disfavouring warmer SEDs even at low redshifts, partly due to confusion in the submm limiting their depth and partly due to the fact that submm wavelengths probe emission from very cold dust at the steep Rayleigh-Jeans part of the SED. 
\item By examining a near complete \textit{IRAS} 60\,$\mu$m sample up to z=0.1, where the survey is sensitive to all SED types with average temperatures in the 17-87\,K range, we confirm that cold ULIRGs are rare in the local Universe. The local ULIRG SEDs have an average peak wavelength of 62$\pm$2\,$\mu$m with an intrinsic dispersion $\sigma$=12\,$\mu$m which translates to a greybody temperature of 42$\pm$2\,K with $\sigma$=8\,K. In comparison, for higher redshift \textit{Spitzer}/MIPS, BLAST and SCUBA samples, we calculate $\left\langle \lambda_{\rm peak}\right\rangle$\,=99$\pm$6\,$\mu$m with an intrinsic dispersion $\sigma$=31\,$\mu$m, 86$\pm$4 with $\sigma$=15\,$\mu$m and 91$\pm$4\,$\mu$m with $\sigma$=11\,$\mu$m ($\left\langle T \right\rangle$\,=26$\pm$2\,K with $\sigma$=8\,K, 29$\pm$1\,K with $\sigma$=5\,K and 29$\pm$1\,K with $\sigma$=3\,K) respectively, indicative of an increase in the relative number of cold galaxies at high redshift. 
\item The local L-T relation which describes luminous objects as warmer than their less luminous counterparts, does not apply to ULIRGs in the distant Universe, as the scatter in the temperature distribution of high redshift ULIRGs is much higher than the ULIRG temperatures predicted by the L-T relation. Only 16 per cent of the MIPS sample falls within 1\,$\sigma$ of the L-T relation, compared to 70 per cent of the local ULIRG sample.
\item The discrepancies between local and high redshift samples, cannot be reconciled by assuming that the local L-T relation evolves with redshift in the same way as the knee of the IR luminosity function, i.e. by (1+z)$^3$. Instead, we find that these differences can adequately be described by the Seymour et al. (2010) results, which show that evolution of the infrared luminosity function is temperature dependent, with cold galaxies evolving at a faster rate, implying that their relative number will decrease with decreasing redshift. 

\end{enumerate}

\section*{Acknowledgments}

MS is grateful for UCL/MSSL support. Thanks to Simon Dye for his help with the BLAST data.

\bibliographystyle{mn2e}
\bibliography{references}

\appendix

\section{The relation between colour and SED peak wavelength}
\label{appendixA}

\begin{figure}
\epsfig{file=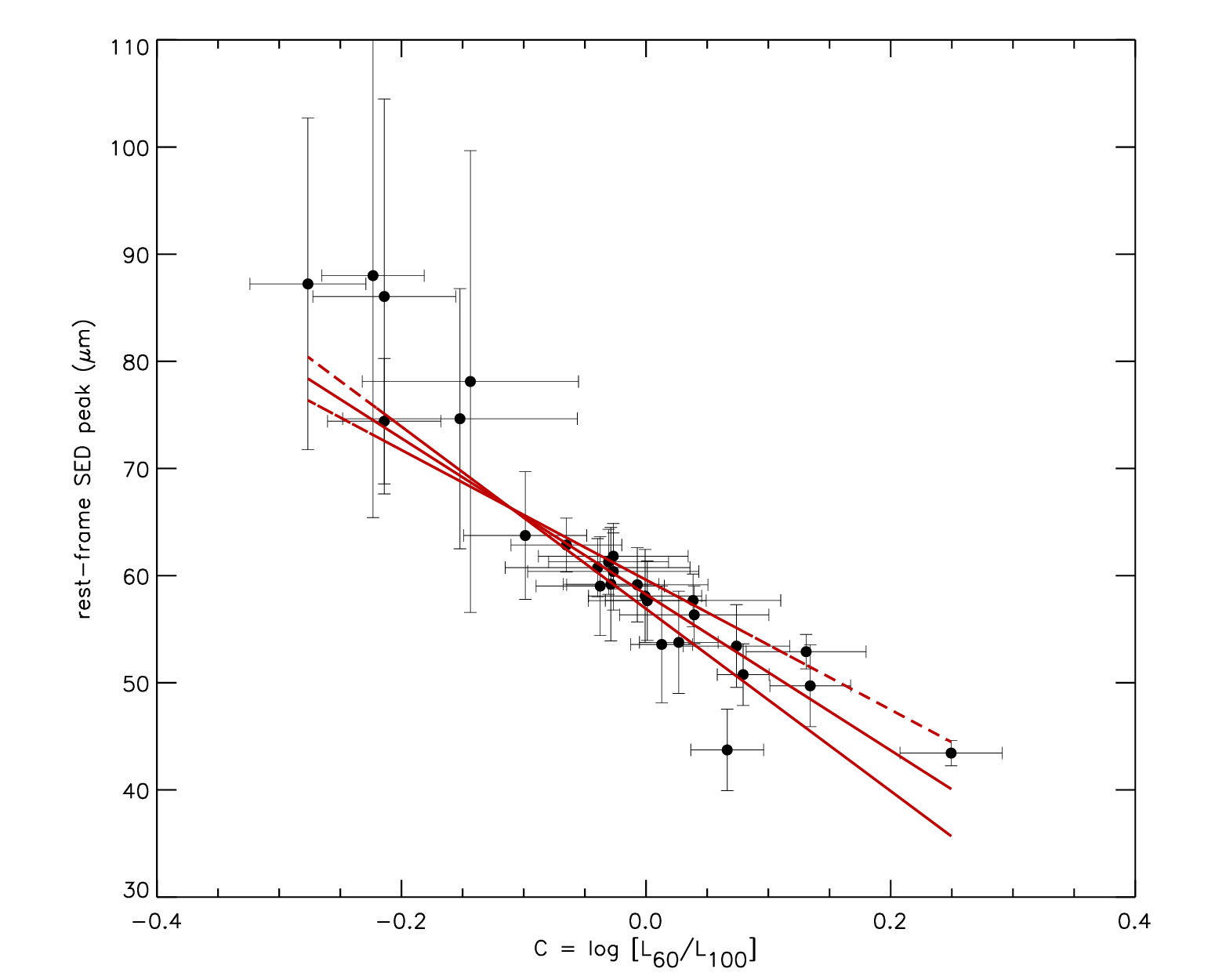,width=8.5cm}
\caption{ SED peak wavelength versus rest-frame colour C [=log\, L$_{60}$/L$_{100}$ where L$_{\lambda}$ is in W/Hz] for the 27 local IIFSCz ULIRGs. The solid and dotted lines are equation \ref{appendix_equation} (and corresponding errors). }
\label{fig:fig9}
\end{figure}

In order to use the CHA09 L-T relation, which uses colour, instead of SED peak wavelength, we establish a conversion between the two quantities. Using our fitted SED templates, we calculate rest-frame L$_{60}$/L$_{100}$ colour (C) for the 27 local IIFSCz ULIRGs (where L$_{\lambda}$ is in W/Hz for direct comparison with the C [= log (f$_{60}$/f$_{100}$) where f$_{\lambda}$ is in Jy] definition used in CHA09). As expected, C and $\lambda_{\rm peak}$  follow a one to one relationship which can be fit with the following equation (using a biweight estimator): 
\begin{equation}
\lambda_{\rm peak}=58.25(\pm 1.35) -72.82(\pm12.2)\,C 
\label{appendix_equation}
\end{equation}
where C=log (L$_{60}$/L$_{100}$), shown in figure \ref{fig:fig9}.

\label{lastpage}

\end{document}